\documentclass[journal]{vgtc}                     

\onlineid{1765}

\vgtccategory{Research}

\vgtcpapertype{algorithm/technique}

\title{Hybrelighter: Combining Deep Anisotropic Diffusion and Scene Reconstruction for On-device Real-time Relighting in Mixed Reality}

\author{%
  Hanwen Zhao,
  John Akers,
  Baback Elmieh,
  Ira Kemelmacher-Shlizerman
}

\authorfooter{
  \item
    Hanwen Zhao, University of Washington, Seattle, Washington, United States, hzhao5@uw.edu\\
    John Akers, UW Reality Lab, University of Washington, Seattle, Washington, United States\\
    Baback Elmieh, Computer Science, University of Washington, Seattle, Washington, United States, baback@cs.washington.edu\\
    Prof Ira Kemelmacher-Shlizerman, University of Washington , Seattle , Washington, United States
}

\abstract{%
  Mixed Reality scene relighting, where virtual changes to lighting conditions realistically interact with physical objects, producing authentic illumination and shadows, can be used in a variety of applications. One such application in real estate could be visualizing a room at different times of day and placing virtual light fixtures. Existing deep learning-based relighting techniques typically exceed the real-time performance capabilities of current MR devices. On the other hand, scene understanding methods, such as on-device scene reconstruction, often yield inaccurate results due to scanning limitations, in turn affecting relighting quality. Finally, simpler 2D image filter-based approaches cannot represent complex geometry and shadows. We introduce a novel method to integrate image segmentation, with lighting propagation via anisotropic diffusion on top of basic scene understanding, and the computational simplicity of filter-based techniques. Our approach corrects on-device scanning inaccuracies, delivering visually appealing and accurate relighting effects in real-time on edge devices, achieving speeds as high as 100 fps. We show a direct comparison between our method and the industry standard, and present a practical demonstration of our method in the aforementioned real estate example.
}

\keywords{Relighting, Mixed Reality, Computer Vision}

\teaser{
  \centering
  \includegraphics[width=\linewidth, alt={A view of a city with buildings peeking out of the clouds.}]{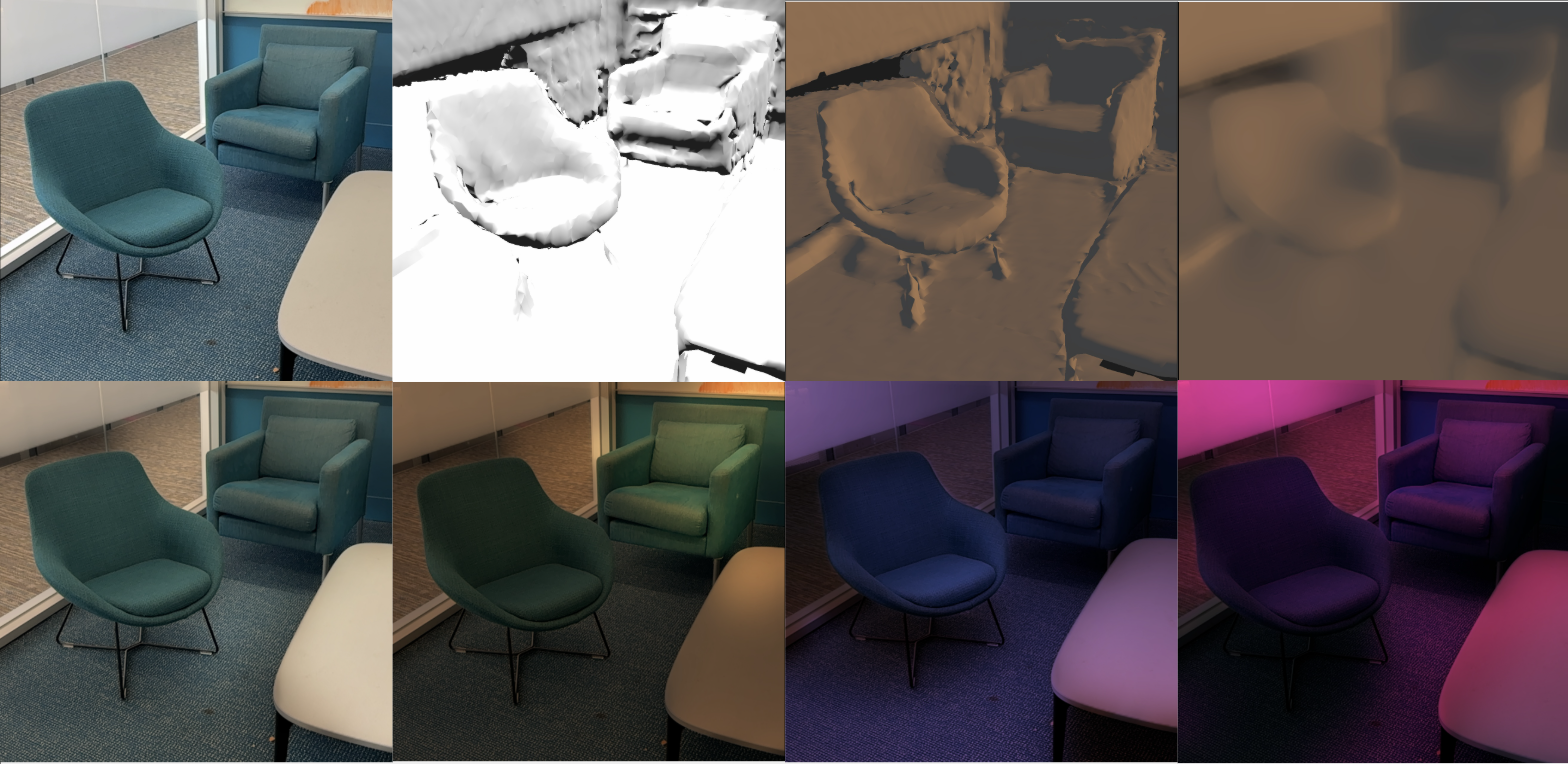}
  \caption{%
    Utilize the scene reconstruction capability and anisotropic diffusion (row 1) for real-time relighting of a room in various lighting conditions in mixed reality (row 2). %
  }
  \label{fig:teaser}
}

\renewcommand{\manuscriptnotetxt}{}

\graphicspath{{figs/}{figures/}{pictures/}{images/}{./}} 

\usepackage{tabu}                      
\usepackage{booktabs}                  
\usepackage{lipsum}                    
\usepackage{mwe}                       

\usepackage{mathptmx}                  
\usepackage{amsmath}
\usepackage{amssymb}

\begin{document}


\firstsection{Introduction}

\maketitle

Recent advancements in Mixed Reality (MR) technologies have significantly enhanced the capabilities of headsets, enabling them not only to visualize but also to comprehend real-world environments through passthrough cameras. This capability allows users to interact with their environments through a controllable virtual rendering layer, opening opportunities for various applications, notably scene relighting. By intelligently manipulating camera data, we can effectively alter the perceived lighting conditions of the environment, offering realistic simulations of diverse lighting scenarios without physically altering actual light sources.

Scene relighting has numerous practical applications, including immersive integration of virtual objects with real-world scenes, creating visually compelling storytelling effects, or simulating environments under different lighting conditions for planning and visualization.

Simple 2D filter-based relighting methods, such as adjustments to image temperature, contrast, or exposure, can be easily implemented with various image editing software packages. More advanced techniques involve masking-based image processing, which allows localized illumination changes, like simulating a spotlight effect. These learning-free approaches are highly efficient, stable, and predictable but suffer from limitations due to the absence of 3D geometric context, particularly when accurately representing complex shadows and occlusion interactions.

In contrast, sophisticated 3D-based relighting methods utilize scene geometry derived from real-time reconstruction to achieve more accurate lighting simulations. These techniques leverage traditional rendering methods, such as rasterization or ray tracing, enabling more realistic representation of shadows and illumination interactions. However, current on-device scene reconstruction, typically performed using built-in lidar or depth sensors on MR headsets, often results in simplified meshes due to computational constraints and limited fidelity, adversely affecting the quality of relighting.

More recent developments in deep learning have paved the way for highly detailed and realistic relighting solutions. Deep neural networks can infer intrinsic scene properties such as surface normals, roughness, albedo, and reflectance, greatly improving the accuracy and visual realism of relighting effects. However, these deep learning methods are typically computationally intensive, limiting their real-time applicability on edge devices such as MR headsets.

In this paper, we propose a novel hybrid method that integrates the computational efficiency of filter-based approaches with the deeper semantic understanding provided by deep learning-guided anisotropic diffusion. Our approach specifically addresses the inaccuracies present in simplified meshes generated by real-time scene reconstruction. By leveraging RGB images captured from passthrough cameras, we employ a deep learning-based feature extractor to accurately identify high-frequency details and object boundaries. Subsequently, we apply anisotropic diffusion iteratively, guided by these learned features, to smoothly interpolate shading within object boundaries while maintaining sharp, precise edges.

Recognizing the computational demands of conventional anisotropic diffusion, which typically require numerous iterations to achieve equilibrium, we introduce a cascaded diffusion strategy. This method capitalizes on the rapid propagation of gradients at lower image resolutions, progressively refining edge details through fewer iterations, significantly enhancing runtime efficiency.
\begin{figure}[tb]
  \centering 
  \includegraphics[width=\columnwidth, alt={An image showing how mesh-aware image filters can be used to effectively relight an image using simple image compositions such as multiplication.}]{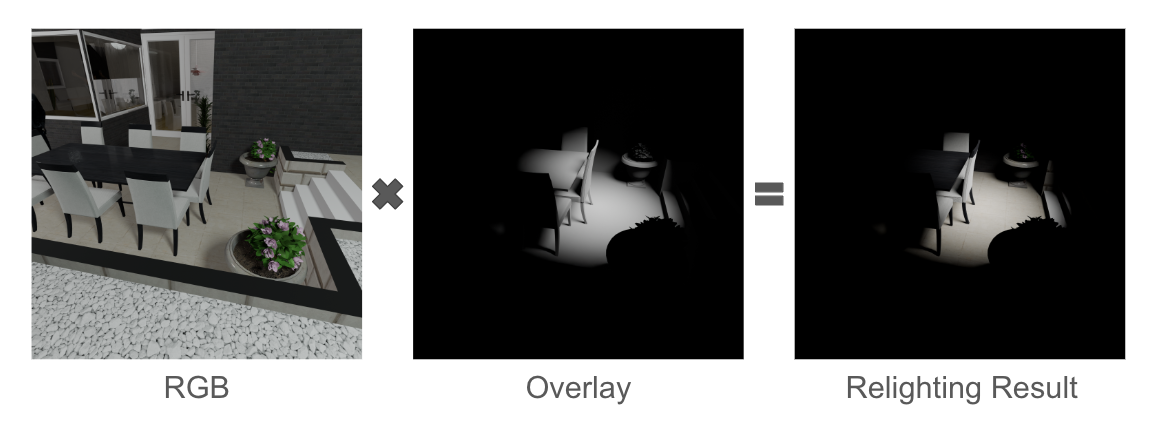}
  \caption{%
  	An example of a synthetic scene being relit by a spotlight image filter rendered on an untextured version of the mesh data.
  }
  \label{fig:vis_papers}
\end{figure}
\begin{figure}[tb]
  \centering 
  \includegraphics[width=\columnwidth, alt={An image showing how real mesh acquired directly from mixed reality devices is suboptimal and the final relit image looks jarring.}]{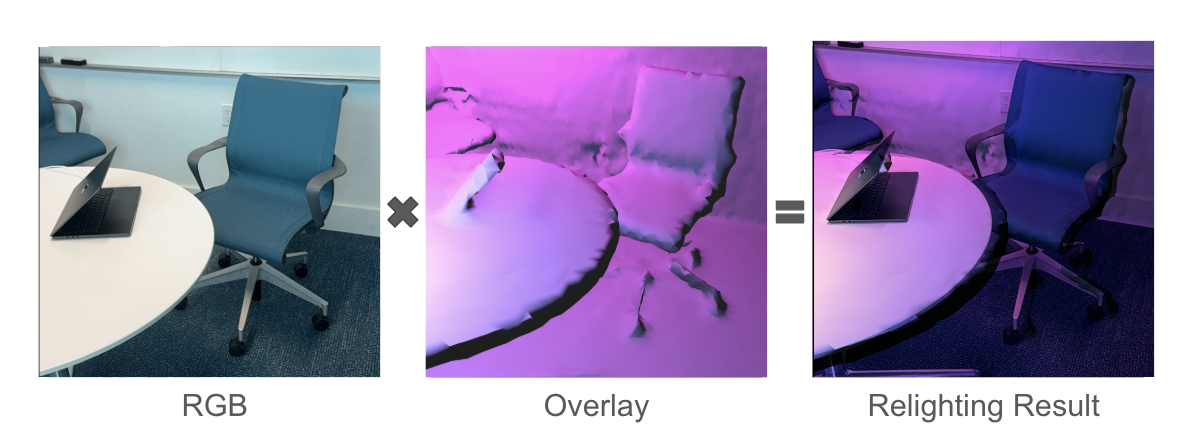}
  \caption{%
  	An example of the mesh quality directly from the LiDAR camera on the iPhone 16 Pro. The errors in such suboptimal mesh can significantly reduce the visual quality of the final relit image, especially around the edges.
  }
  \label{fig:vis_papers}
\end{figure}

In short, the problem we are solving is: How can we relight a dynamically updating scene that is geometrically correct, and runs in real-time on device in Mixed Reality? And our primary contributions include:

\begin{itemize}
      \item A hybrid relighting approach that effectively combines deep learning feature extraction with filter-based anisotropic diffusion.
      \item A modified anisotropic diffusion process optimized for fewer iterations, demonstrating that existing deep diffusion model architectures can be effectively adapted to run in real time on edge devices.
      \item A practical demonstration and evaluation of our method on commercially available MR devices. We present how relighting can be practically applied in a real estate tour scenario.
\end{itemize}

\section{Related Work}

\textbf{Scene Reconstruction based Relighting Approaches} \\
Scene relighting techniques have historically relied on high-quality reconstructions as a basis for realistic lighting simulations. Early research demonstrated the potential of mesh-based methods to produce convincing relit images given sufficiently detailed geometry \cite{article}. However, due to computational constraints, MR headsets typically employ mesh decimation strategies, resulting in reduced fidelity and diminished relighting quality. To address this limitation, several studies have proposed methods to incrementally refine meshes dynamically and efficiently. Hybrid voxel-octree fusion techniques \cite{liu2024hvo}, and the Large Reconstruction Model (LRM) \cite{wei2024meshlrm}, demonstrate that high-quality meshes can be rapidly generated even from sparse input views. Additionally, recent advancements in real-time dense scene reconstruction can achieve high-fidelity meshes at approximately 20 frames per second (fps) on powerful GPUs, although these methods remain computationally prohibitive for edge-device deployment \cite{slam3r}.

Neural rendering methods, including Gaussian splatting and Neural Radiance Fields (NeRF), have also been explored for their impressive capabilities in large-scale scene relighting \cite{liu2024citygaussian} \cite{bi2024rgs} \cite{zeng2023nrhints}. Despite achieving visually compelling results, these methods require extensive pre-training, ranging from several minutes to half an hour, thus limiting their practical usability in interactive MR scenarios.

Other recent approaches seek to infer 3D scene information without explicitly generating 3D representations. Advances in depth estimation have enabled consistent depth image generation from extended video sequences, achieving high-performance processing speeds of approximately 9 ms per frame (over 100 fps) on high-end discrete GPUs \cite{video_depth_anything}. Additionally, depth estimation methods tailored for edge devices have been developed, reaching speeds up to 300 fps on embedded Nvidia hardware, albeit with compromised image quality and temporal inconsistency \cite{rtmonodepth}.

An approximation of surface normals can be efficiently derived using camera intrinsics and pixel-wise cross-products; however, accurately capturing complex occlusion information necessary for realistic shadow rendering introduces additional computational overhead, oftentimes requiring another learning-based model running on top of it as explored in the work by Yang et al. \cite{yang2021multi}. The quality of relighting thus significantly depends not only on global depth consistency across frames but also on local depth accuracy. To address this, joint prediction frameworks for depth and surface normals have emerged, demonstrating superior quality but sacrificing real-time performance on edge computing devices \cite{hu2024metric3d}.

\textbf{Image-based Direct Relighting Methods} 

Image-based relighting has been extensively investigated within computer vision literature. Traditional techniques typically involve decomposing images into intrinsic components, such as albedo, shading, and lighting, often requiring multiple neural network layers for accurate extraction \cite{Yi_2023_CVPR} \cite{Kim_2024_CVPR}. More recent research leverages generative models capable of directly synthesizing relit images conditioned upon new lighting parameters, achieving high-quality results with simpler architectures. These generative models demonstrate remarkable flexibility and visual realism \cite{ponglertnapakorn2023difareli} \cite{StyLitGAN} \cite{jin2024neural_gaffer}. Although explicit runtime evaluations are not always provided, related studies on diffusion-based generative methods indicate that one-step diffusion models can produce relit images as rapidly as 9 ms per frame on high-end GPUs \cite{song2024sdxs}. However, all these methods are tailored for static images, without considering temporal information which makes maintaining consistent illumination across video sequences in real-time a challenge requiring further investigation.

\textbf{Guided Depth and Point Cloud Enhancement}

Working with low-fidelity data from depth sensors and lidar scanners is a significant challenge in many fields, including MR, autonomous driving, and robotics. Thus having high-quality depth and point cloud data are crucial for accurate scene understanding and effective relighting. Recent approaches have leveraged concurrent RGB imagery to enhance \cite{Metzger_2023_CVPR} \cite{conde2024compresseddepthmapsuperresolution} \cite{zhao2023sphericalspacefeaturedecomposition} \cite{sun2022consistent} and complete sparse or imperfect depth information \cite{Shi_2024_CVPR} \cite{DFU_CVPR_2024}, exploiting complementary information between modalities. Similar approaches can also be applied on point-cloud completion \cite{liu2024meshformer} \cite{li2024dapointrdomainadaptivepoint}. Studies on guided depth super-resolution and depth completion indicate that real-time performance (up to 50 fps) is achievable, highlighting the potential of these methods for practical MR deployment \cite{9357967}.

Our proposed method draws inspiration from this body of work. Specifically, we combine RGB guidance and suboptimal mesh data to significantly improve scene relighting quality. Our approach employs anisotropic diffusion guided by learned RGB features, effectively translating established depth super-resolution methods into the context of real-time MR relighting.

\section{Methods}

\begin{figure}[tb]
  \centering 
  \includegraphics[width=\columnwidth, alt={An image of our relighting pipeline in high level.}]{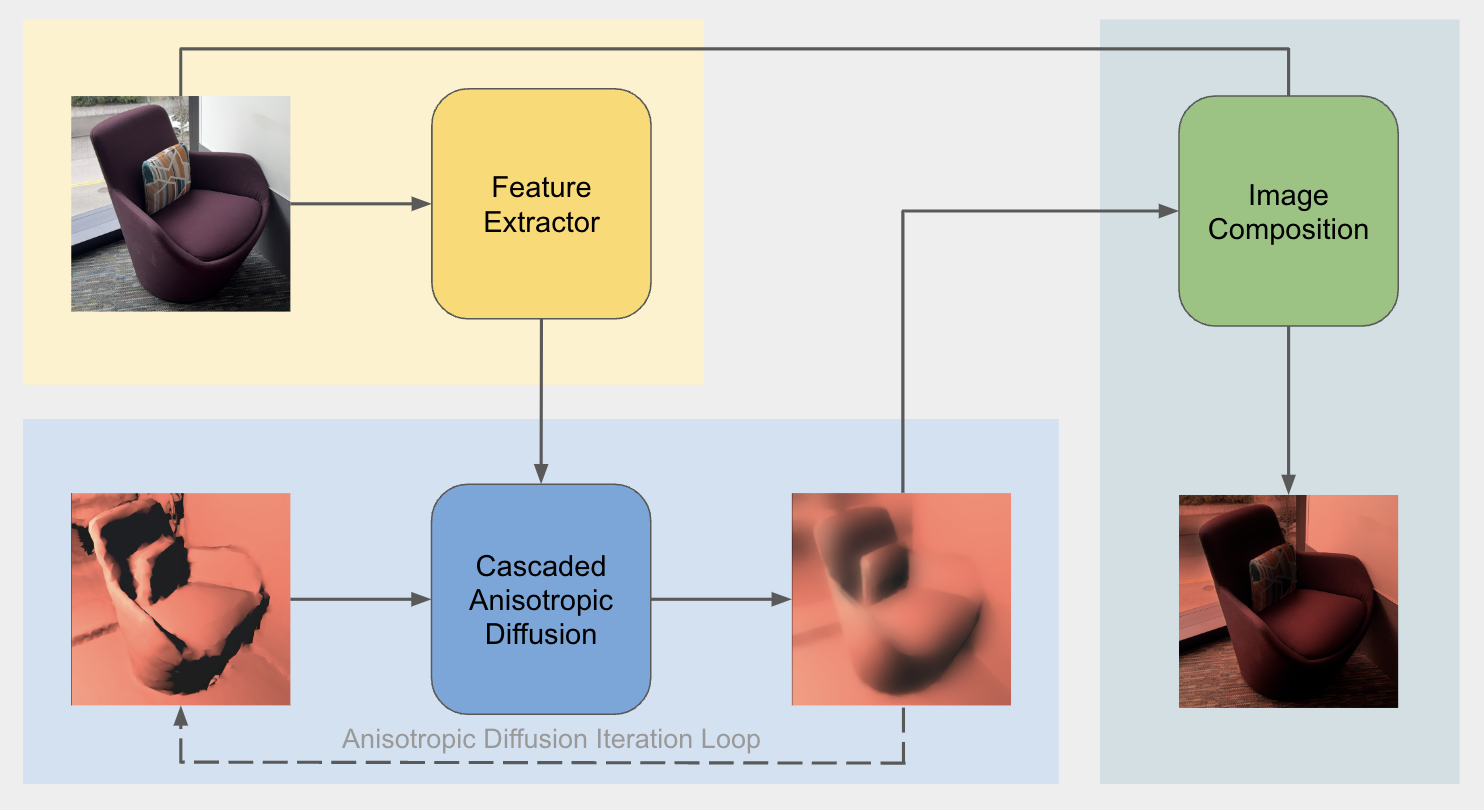}
  \caption{%
  	Our relighting pipeline takes in one RGB camera image and one RGB relight image. High-level edge information extracted from the RGB camera image is used as the guidance for anisotropic diffusion running on the relight image. The refined output is composited with the original camera RGB to produce the final relit image.
  }
  \label{fig:vis_papers}
\end{figure}

Our proposed relighting method integrates a mesh-aware, filter-based approach with guided anisotropic diffusion, delivering high-quality, real-time relighting for mixed reality (MR) scenarios on edge devices. This method combines the speed and efficiency of GPU-based filtering with the semantic precision of deep learning-based anisotropic diffusion, effectively overcoming the visual quality limitations introduced by simplified mesh reconstruction typically found in MR hardware. The following sections detail each component of our pipeline, including mesh-aware filtering, guided anisotropic diffusion, cascaded diffusion strategies for enhanced computational efficiency, shadow rendering adjustments, and transferable training methodologies derived from existing depth super-resolution techniques.
\subsection{Mesh-aware Filter-based Relighting}
To leverage the computational speed of simple 2D-filters we utilize a mesh-aware filtering pipeline that integrates geometric awareness into traditional image processing. Initially, we reconstruct the scene geometry using real-time mesh reconstruction techniques typically available on MR devices. Within a 3D rendering engine, we introduce virtual lights and render the mesh using standard shading algorithms. The resulting shaded mesh acts as an image-space filter that is precisely aligned with the RGB camera frames through consistent camera intrinsic parameters. The final relit image is obtained by compositing this rendered mesh with the original RGB frame via image multiplication.

Thus, it can be defined as the following: we are given a rendered relight image $ R \in \mathbb{R} ^ {H\times W \times 3} $ serving as the filter acquired directly from the original mesh. And we are given an RGB camera frame $C \in \mathbb{R} ^ {H \times W \times 3}$ as the passthrough. The output which is our relit image which can be simply defined as 
\begin{equation}
S = R \odot C
\end{equation}
where \( \odot \) denotes element-wise (per-pixel) multiplication.
While this approach provides efficiency and real-time performance, it depends significantly on the accuracy and fidelity of the reconstructed mesh. Due to the computational limitations inherent in MR hardware, meshes are often simplified, resulting in degraded visual quality and inaccurate shading outcomes. 
\subsection{Guided Anisotropic Diffusion for Relighting Correction}
Our key insight is that the mesh inaccuracies that affect shading quality most severely occur at 3D depth discontinuities which correspond to edge boundaries in 2D images. To solve for this, we employ anisotropic diffusion, an edge-aware filtering technique known for effectively smoothing interior regions of objects while preserving edges. Specifically, we guide anisotropic diffusion using edge information extracted by a deep learning-based feature extractor trained to identify critical scene details and object boundaries. By using learned features rather than simple RGB edges, we enhance diffusion precision, ensuring the gradient propagation remains confined within object boundaries.

Drawing inspiration from guided depth super-resolution research \cite{Metzger_2023_CVPR}, we adapt the anisotropic diffusion approach from the depth domain to the relighting scenario. The guided anisotropic diffusion can be mathematically described using the formulation introduced by Metzger et al. \cite{Metzger_2023_CVPR}. The prediction of a pixel's value at iteration \(t\) at location \(p\) is expressed as follows:
\begin{equation}
\hat{y}_t^p = y_{t-1}^p + \lambda \cdot \sum_{n \in \mathcal{N}_4^p} \left( y_{t-1}^n - y_{t-1}^p \right) \cdot c(\mathbf{g}^p, \mathbf{g}^n)
\end{equation}
Here, \(y_t^p\) represents the pixel intensity at position \(p\) in the image \(\mathbf{Y}_t\). \(\mathcal{N}_4^p\) refers to the set of four directly adjacent pixels (4‑neighborhood) around pixel \(p\), effectively forming a planar graph across the image. The hyperparameter \(\lambda\) controls the diffusion strength and ensures stability during iterations; when utilizing 4‑neighborhood connectivity, it must satisfy \(\lambda < \frac{1}{4}\).
The diffusion coefficient \(c(\mathbf{g}^p, \mathbf{g}^n)\) is computed based on the similarity between guide features at pixels \(p\) and \(n\), as initially introduced by \cite{Metzger_2023_CVPR}:
\begin{equation}
c(\mathbf{g}^p, \mathbf{g}^n) = \frac{\kappa^2}{\kappa^2 + \lVert \mathbf{g}^p - \mathbf{g}^n \rVert_2^2}
\end{equation}
The hyperparameter \(\kappa\) controls sensitivity to feature gradients within the guidance image \(\mathbf{G}\), with higher values allowing smoother diffusion across larger differences. The symmetry of this coefficient function ensures \(c(\mathbf{g}^p, \mathbf{g}^n) = c(\mathbf{g}^n, \mathbf{g}^p)\).

While conventional anisotropic diffusion methods directly use the current image state \(\mathbf{Y}_{t-1}\) as guidance, the method we adopt explicitly incorporates separate guidance data \(\mathbf{G}\). This distinction is particularly valuable in relighting tasks. Unlike depth maps, relit images often contain inaccuracies due to simplifications in the initial mesh representation. Thus, our method addresses these inherent inaccuracies explicitly, optimizing the diffusion process to yield visually consistent results in fewer iterations.

\begin{figure}[t]
  \centering 
  \includegraphics[width=\linewidth, alt={An image showing the effect of guided anisotropic diffusion on Gaussian noise.}]{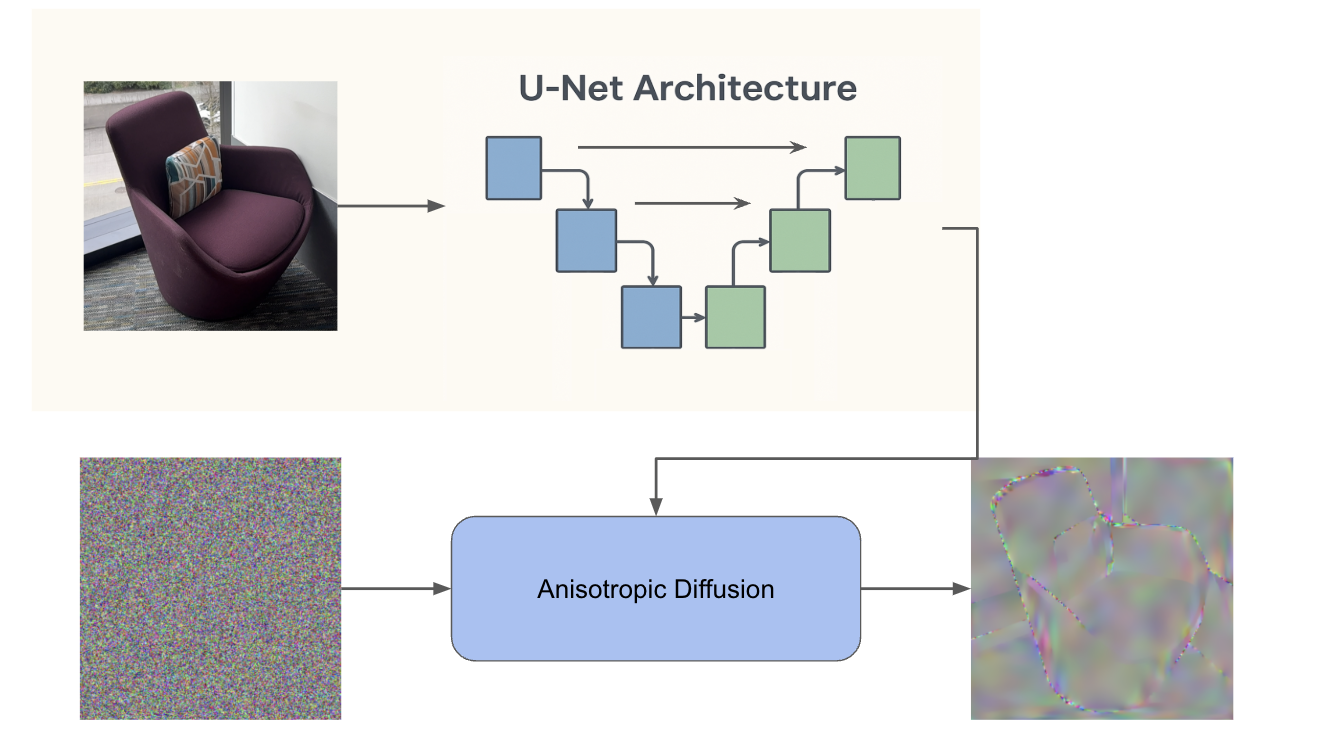}
  \caption{%
  	The effect of the anisotropic diffusion coefficients provided by the RGB camera guidance image. The input for anisotropic diffusion is pure Gaussian noise. The output is scaled up for better image clarity. %
  }
  \label{fig:vis_papers}
\end{figure}

\begin{figure}[h]
  \centering 
  \includegraphics[width=\linewidth, alt={Two side-by-side comparison images showing how guided anisotropic diffusion can correct scanning errors while maintaining high edge consistency.}]{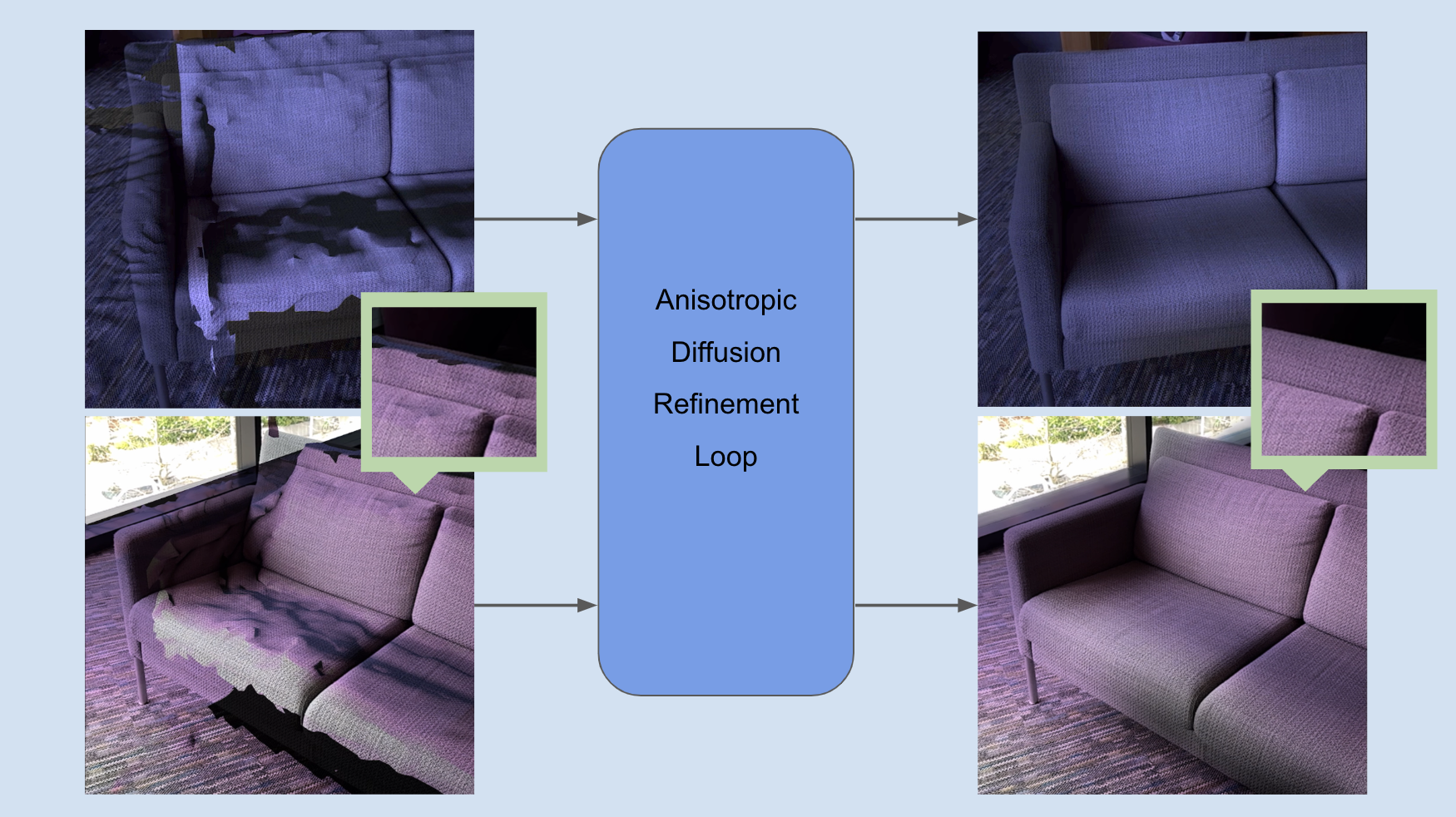}
  \caption{%
  	Edge-aware pixel-color diffusion process demonstrates strong capabilities in fixing the errors caused by inaccuracies in the scanning process. We show how this approach can produce smooth shading in each object while maintaining high edge consistency. %
  }
  \label{fig:vis_papers}
\end{figure}

\subsection{Cascaded Anisotropic Diffusion for Enhanced Efficiency}
Recognizing the computational burden of conventional anisotropic diffusion, we introduce a cascaded diffusion strategy operating at multiple resolutions.
\begin{align}
&\textbf{(1) Initialize coarse level:} \nonumber \\
&\quad \mathbf{Y}_0^{(1)} = \downarrow^{L-1}(\mathbf{Y}_0), \quad \mathbf{G}^{(1)} = \downarrow^{L-1}(\mathbf{G}) \nonumber \\[6pt]
&\textbf{(2) Coarse-to-fine diffusion:} \nonumber \\
&\quad \text{For } l = 1 \text{ to } L: \nonumber \\
&\qquad \mathbf{C}^{(l)} = \text{MinPool}\left(c(\mathbf{G}^{(l)})\right) \nonumber \\
&\qquad \mathbf{Y}_t^{(l)} = \mathcal{D}\left(\mathbf{Y}_{t-1}^{(l)}, \mathbf{C}^{(l)}\right) \nonumber \\[6pt]
&\textbf{(3) Upsample and refine:} \nonumber \\
&\qquad \text{If } l < L: \quad \mathbf{Y}_t^{(l+1)} = \uparrow\left(\mathbf{Y}_t^{(l)}\right) \nonumber \\[6pt]
&\textbf{(4) Final relit image:} \nonumber \\
&\quad \mathbf{S} = \mathbf{Y}_t^{(L)}
\end{align}
In this formulation, let \( L \) be the number of resolution levels, where \( l = 1 \) denotes the coarsest scale and \( l = L \) corresponds to the original resolution. The image \( \mathbf{Y}_t^{(l)} \) and guide \( \mathbf{G}^{(l)} \) are derived by downsampling the original inputs \( \mathbf{Y}_0 \) and \( \mathbf{G} \) respectively. At each level, we compute diffusion coefficients \( \mathbf{C}^{(l)} \) using a min pooling strategy, ensuring edge preservation across pooling grids. The anisotropic diffusion operator \( \mathcal{D} \) is then applied using these coefficients. The diffused image is upsampled to the next finer level, where additional refinement iterations are performed. This coarse-to-fine cascade continues until the finest resolution is reached, yielding the final relit image \( \mathbf{S} \). This approach significantly reduces computational overhead while preserving critical high-frequency details needed for sharp visual relighting.
By initially performing diffusion at reduced image resolutions, we significantly expedite gradient propagation across large pixel areas. We employ min pooling for downsampling diffusion coefficients, ensuring edge integrity by treating any edge pixel as an edge across the entire pooling grid, thereby preventing unintended color blending.

Subsequently, we progressively upscale the image back to its original resolution, applying additional, targeted diffusion iterations at each resolution level to restore detailed edge information lost during downsampling. This cascading approach significantly reduces computational requirements, enabling real-time performance while preserving critical high-frequency details necessary for visually sharp relighting.

\begin{figure}[h]
  \centering 
  \includegraphics[width=\linewidth, alt={A figure showing cascaded anisotropic diffusion.}]{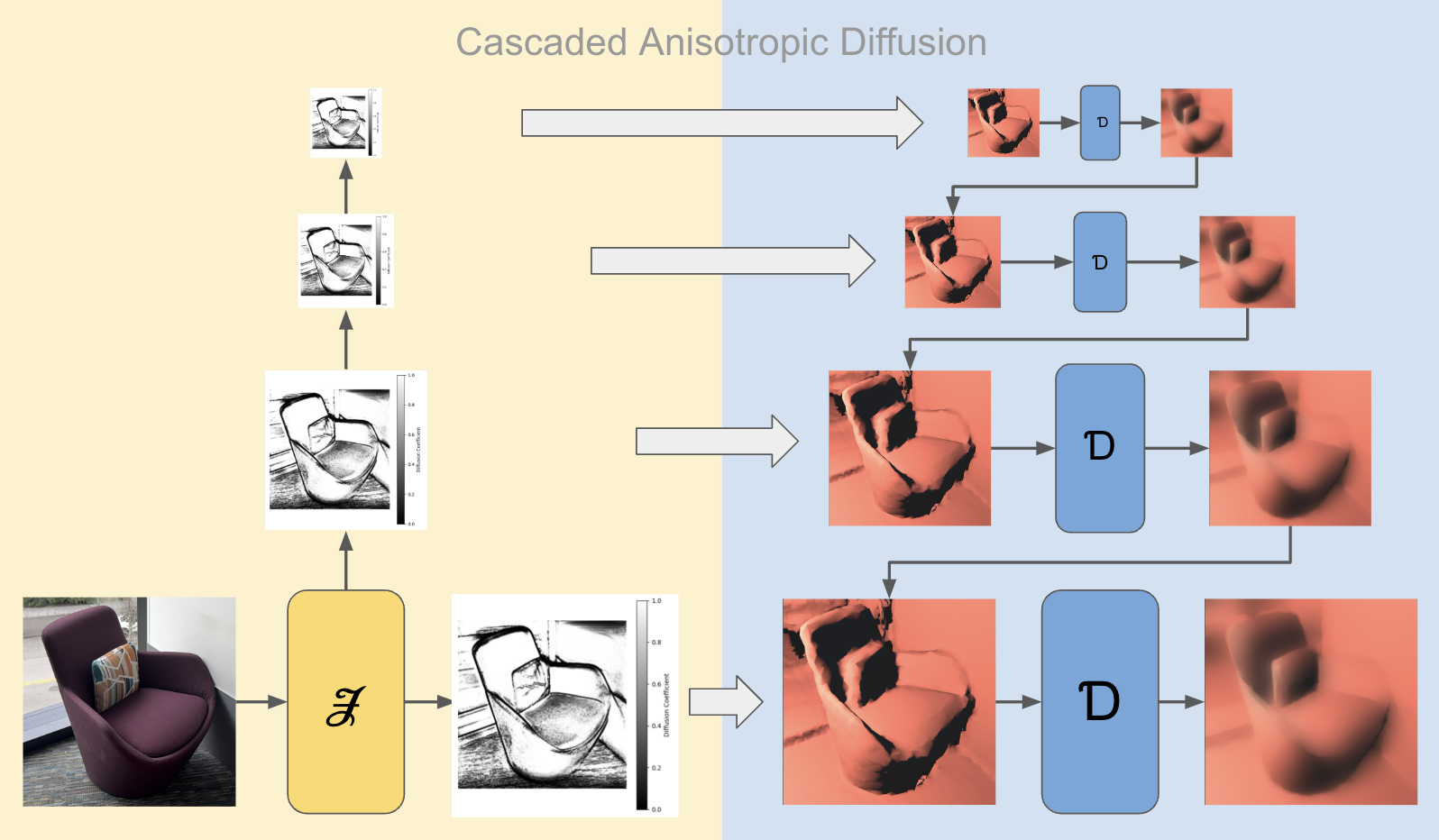}
  \caption{%
  	Our cascaded anisotropic diffusion pipeline. We run the diffusion process at lower resolution and we chain the results together by up-sampling the previous outputs and further refining image quality. %
  }
  \label{fig:vis_papers}
\end{figure}

\subsection{Adjusted Anisotropic Diffusion for Shadow Rendering}
\begin{figure}[tb]
  \centering 
  \includegraphics[width=\linewidth, alt={A figure showing the separate diffusion pass for shadows to ensure their shape integrity.}]{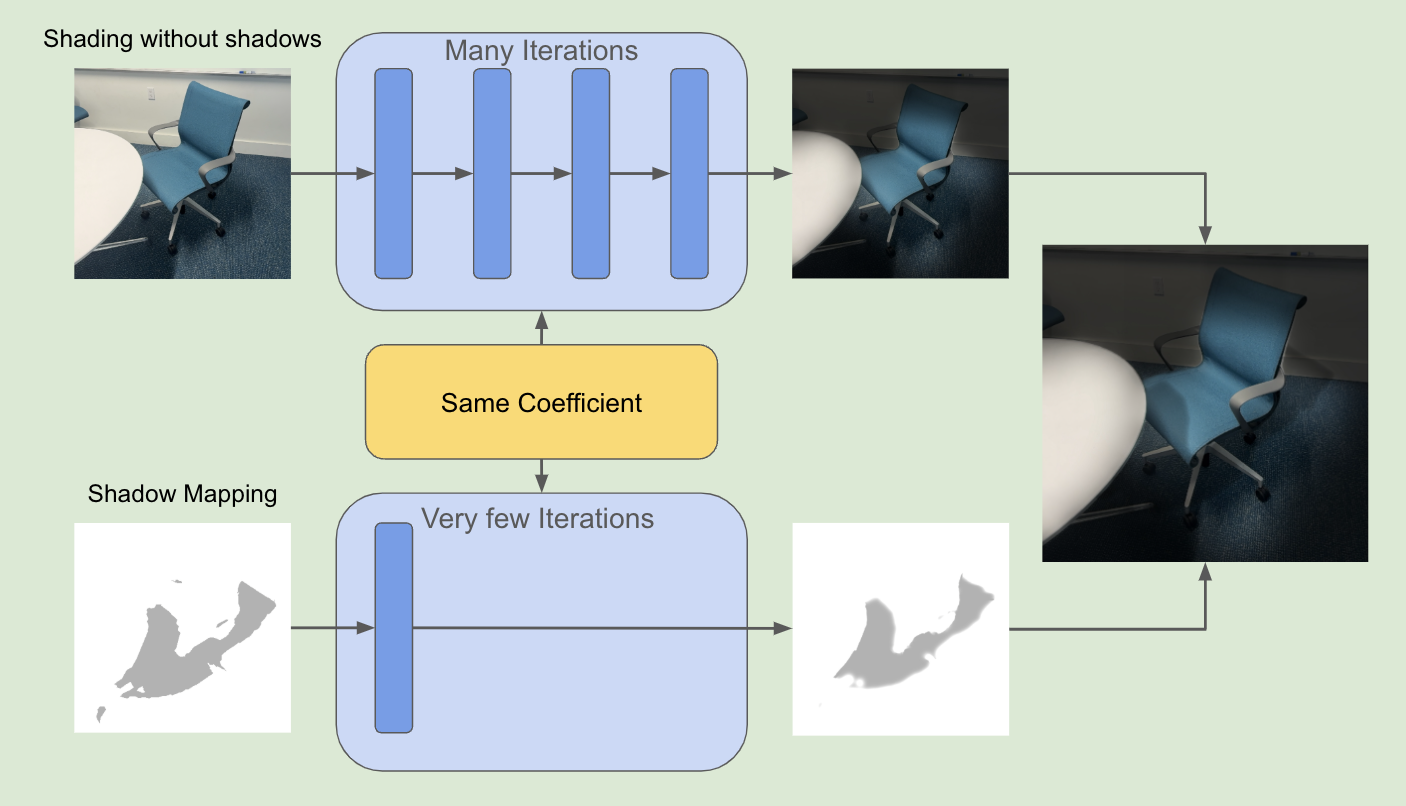}
  \caption{%
  	Shadows can be computed in a separate pass using traditional shadow mapping techniques together with only a few of anisotropic diffusion iterations to ensure their overall shapes are not destructed by the diffusion process.%
  }
  \label{fig:shadowpass}
\end{figure}

An important aspect of realistic relighting is accurately rendering shadows cast by real objects. Unlike direct shading, shadows' shapes depend on the geometry of casting objects rather than the surfaces onto which they fall. To preserve the accuracy and softness of shadow edges, we implement a separate diffusion pass specifically optimized for shadow rendering, as shown in \cref{fig:shadowpass}. By reducing the number of diffusion iterations and maintaining operation at higher resolutions, we effectively blur shadows to produce soft, realistic effects without compromising shape accuracy. The edge-aware nature of the anisotropic diffusion process ensures that shadow colors remain confined, preventing leakage onto adjacent objects. \\

\subsection{Transferable Training from Depth Super-resolution Models}
The core learnable component of our method is the deep learning-based feature extractor responsible for guiding anisotropic diffusion. Leveraging existing depth super-resolution training frameworks, we train a compact, state-of-the-art model suitable for edge-device deployment. Our training pipeline modifies the input channels from single-channel depth maps to three-channel RGB images, facilitating the direct application of existing depth datasets by artificially randomizing color tints. This approach significantly simplifies the training process, benefiting from widely available depth datasets while addressing the scarcity of annotated relighting datasets.

\begin{figure}[h]
  \centering 
  \includegraphics[width=\linewidth, alt={A figure from \cite{Metzger_2023_CVPR} that visualizes diffusion coefficients for both RGB and deeplearning-based}]{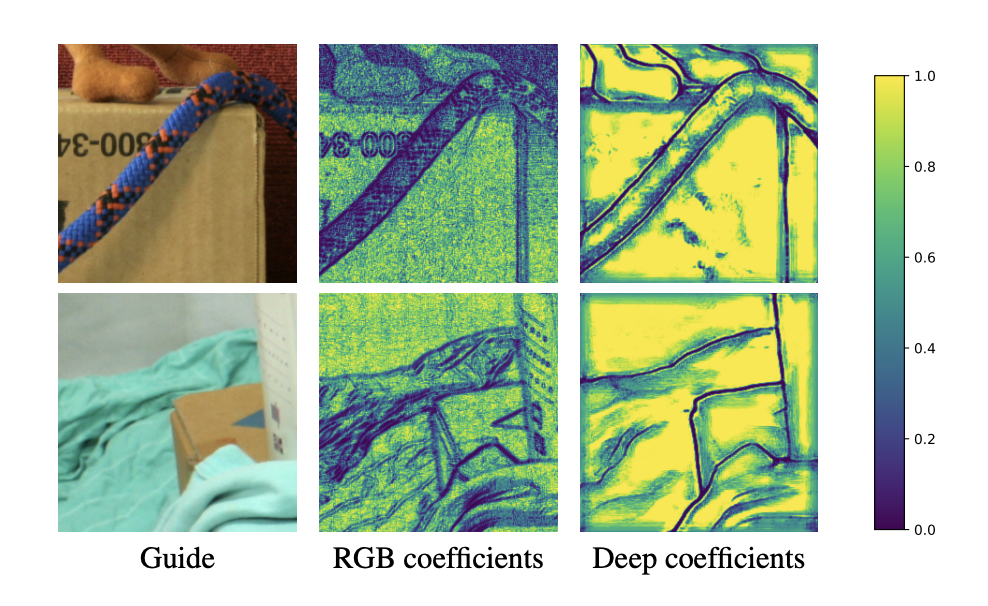}
  \caption{%
        A figure from \cite{Metzger_2023_CVPR}. This figure directly visualizes the values of the coefficients. We observe that they have the exact properties we want for guided anisotropic diffusion. Since this is the only trainable section in the whole pipeline, this shows that any well trained deep feature extractor can be used for any guided anisotropic diffusion tasks.
  }
  \label{fig:orig_paper}
\end{figure}

\section{Experiments}

\subsection{Training Details}
The trainable component within our relighting pipeline is the deep feature extractor used on RGB camera images. An effective feature extractor ideally produces values close to zero at edges and close to one within object regions, facilitating smooth gradient propagation internally while preserving sharp boundaries. We observe this characteristic to be universally beneficial regardless of the specific downstream tasks, as the optimization consistently targets the diffusion coefficients. Consequently, our training procedure is significantly inspired by the approach detailed in Metzger et al. \cite{Metzger_2023_CVPR}.

We use widely accessible datasets from the domain of depth super-resolution for training, specifically the DIML dataset. Our experiments involve training two state-of-the-art mobile-oriented vision models: mobilenetv4\_conv\_small and mobilenetv4\_conv\_medium. Both models employ a U-Net structure from the segmentation\_models\_pytorch package to generate precise feature maps.

A notable departure from the original method presented by Metzger et al. is that our feature extraction exclusively relies on RGB images, omitting the concatenation of lower-resolution depth inputs. This change is essential because the relighting images available to us inherently contain inaccuracies and inconsistencies, thus precluding their use as reliable ground truth even at reduced resolutions. Removing these erroneous inputs ensures that our models do not learn to depend on potentially faulty data.
Another significant innovation introduced by Metzger et al., known as "adjustment step," is crucial for ensuring convergence during training on depth datasets. We also integrate this technique into our training procedure to improve stability and accuracy.

Lastly, we clearly outline our training parameters, highlighting key adjustments from the original configuration. We adopt the same dynamically stepped learning rate as proposed by Metzger et al. but use higher-resolution 512x512 images to ensure enhanced edge fidelity. The depth upsampling scale factor is set at 32. Due to memory constraints during backpropagation and data tracking, we configure the training step as 512 and predictions as 20,000. Both vision models are trained on NVIDIA A100 GPUs, with a batch size of 8 over 12,000 iterations.

\subsection{Numerical Metrics}
We conduct comprehensive evaluations of our method using the ARKitScenes dataset \cite{dehghan2021arkitscenes}, which is collected by Apple using the same LiDAR camera embedded within their mixed reality devices. This dataset comprises RGB camera frames, ARKit-generated meshes identical to those accessible on-device, and high-density point clouds acquired by a Faro laser scanner, providing an ideal benchmark for assessing our relighting approach.

Given that the detailed point cloud data surpasses the real-time processing capabilities of edge devices, we perform offline rendering of both the ARKit-generated meshes and the high-density point clouds using Blender. This ensures alignment consistency between the rendered meshes and the RGB camera frames. Subsequently, we employ a mesh-based filtering strategy in Blender by adding virtual light sources directly onto both meshes and point clouds. The rendered outputs from this step are composited with the corresponding RGB camera frames to generate final relit images, ensuring accurate alignment by utilizing the provided pose and trajectory information.

To assess our method, we introduce three benchmarks, each designed to evaluate distinct aspects of our pipeline:

\textbf{Benchmark 1: Mesh Error Correction}

We first evaluate our method's effectiveness at correcting inaccuracies inherent in the ARKit mesh. In this scenario, we simulate an ideal lighting condition by placing a bright point light centrally within the room. A flawless mesh rendering would yield a grey to white filter, implying low errors in the composited result. Conversely, an imperfect mesh introduces black pixels in the filter, resulting in artifacts in the final composited image. We quantitatively measure performance using both LPIPS \cite{zhang2018perceptual} and Peak Signal to Noise Ratio (PSNR) metrics, comparing:
\begin{itemize}
    \item Original RGB frames and directly composited ARKit mesh-based rendering
    \item Original RGB frames and our refined anisotropic diffusion-enhanced rendering
\end{itemize}
Our results demonstrate that our refined method achieves lower LPIPS and higher PSNR scores, verifying its capability to effectively mitigate artifacts caused by mesh inaccuracies.

\textbf{Benchmark 2: Multi-Lighting Consistency}

Acknowledging that trivial solutions (such as a uniformly white filter) could artificially perform well in Benchmark 1, we introduce a second scenario that mimics realistic relighting conditions more closely. Here, two dimmed, differently-colored point lights are placed within the scene to generate a different lighting environment. Again, we assess the rendered images against original RGB frames using LPIPS and PSNR metrics. Our refined method consistently exhibits lower LPIPS scores compared to direct mesh rendering, emphasizing its ability to retain the visual features of the original scenes under varied lighting configurations.

\textbf{Benchmark 3: Relighting Fidelity}

Lastly, we evaluate our method’s capability to generate visually accurate relighting effects relative to a high-fidelity baseline rendered from dense point cloud data. By comparing:
\begin{itemize}
    \item Our refined method vs. the high-fidelity point cloud rendering
    \item Direct mesh rendering vs. the high-fidelity point cloud rendering
\end{itemize}

Although anisotropic diffusion can introduce slight pixel color differences due to inherent smoothing, as reflected in the PSNR scores, our approach demonstrates strong performance in preserving structural consistency and overall visual realism, as evidenced by lower LPIPS scores. This outcome aligns with our intuition: since the anisotropic diffusion result serves primarily as a filter that preserves edges while smoothing other areas, compositing it with the original, sharp RGB image mitigates potential noise introduced when smoothing was performed. Therefore, the clarity and sharpness of the original RGB image ensure that the final composited output maintains high visual fidelity. 

\begin{figure}[]
  \centering 
  \includegraphics[width=\linewidth, alt={A figure showing comparison between mesh-based and our approach on refining artifacts in ARKit mesh. }]{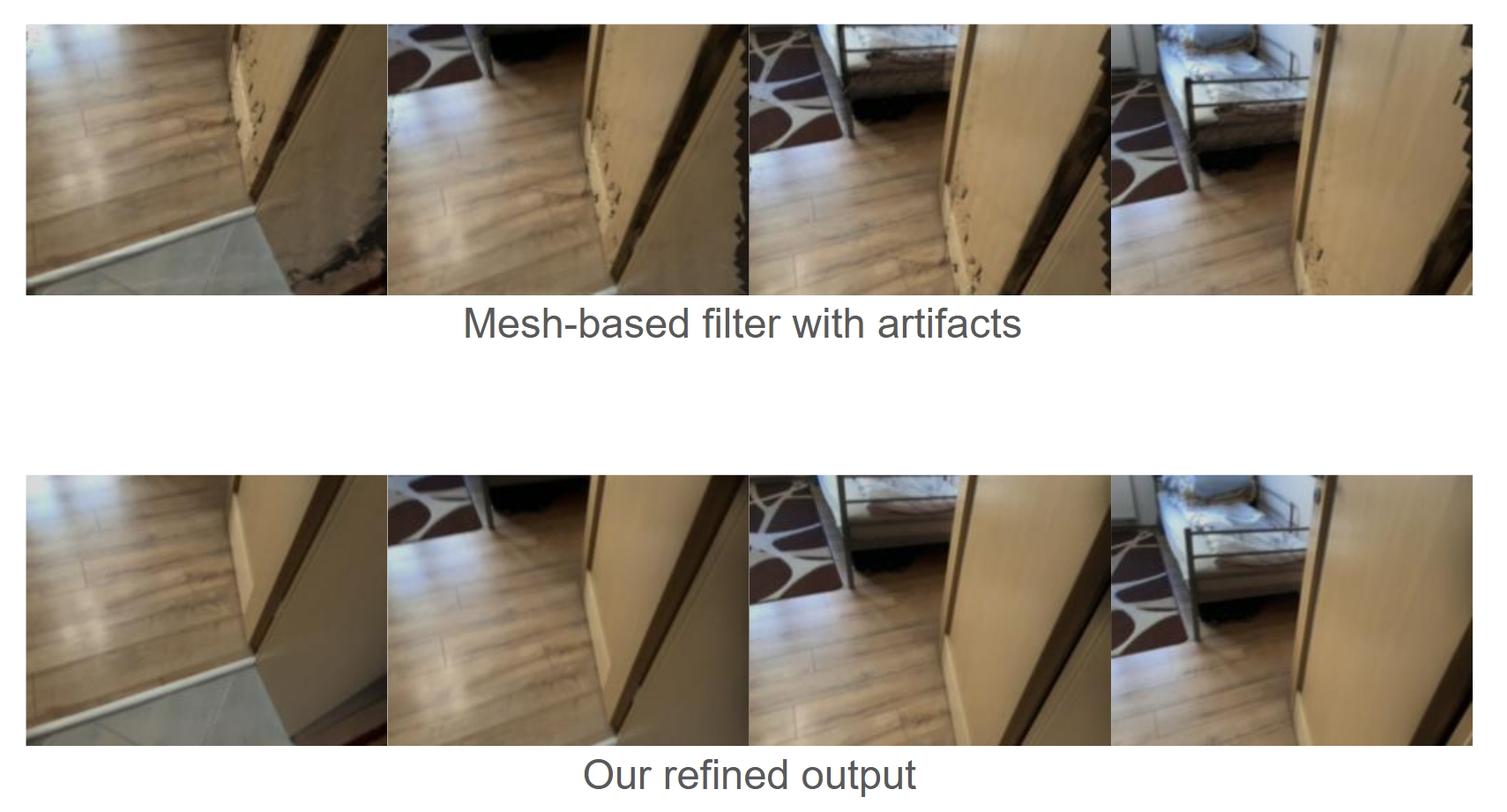}
  \caption{%
        Side by side comparison showing how anisotropic diffusion can correct the artifacts that exist in the mesh directly acquired by ARKit.
  }
  \label{fig:meshRefine}
\end{figure}

\begin{figure}[]
  \centering 
  \includegraphics[width=\linewidth, alt={A figure showing comparison between mesh-based and our approach on relighting.}]{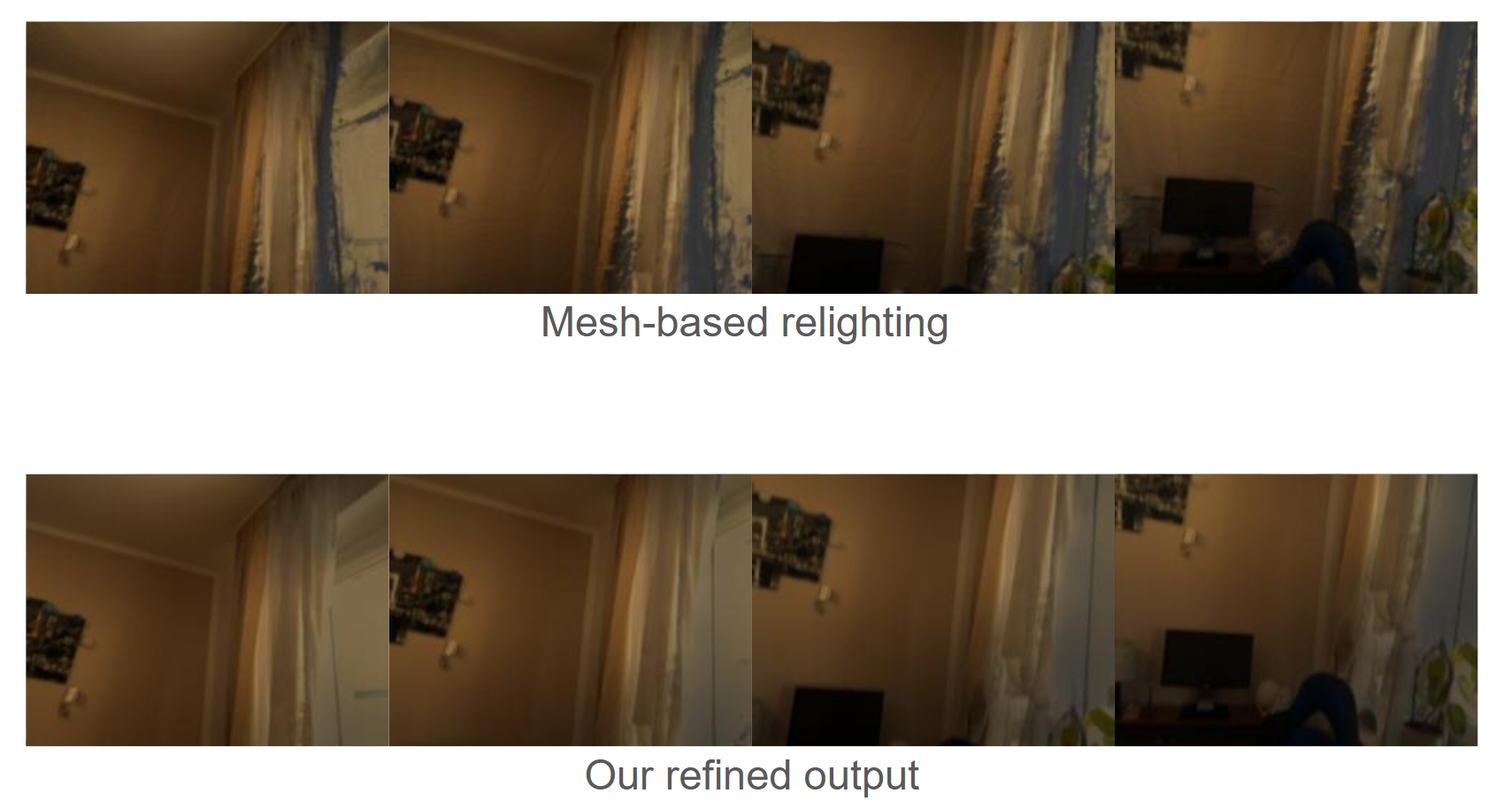}
  \caption{%
        Side by side comparison showing our method and direct mesh-based filter approach on relighting tasks.
  }
  \label{fig:meshRelight2}
\end{figure}

\begin{figure}[]
  \centering 
  \includegraphics[width=\linewidth, alt={A figure comparing results with and without separate shadow pass.}]{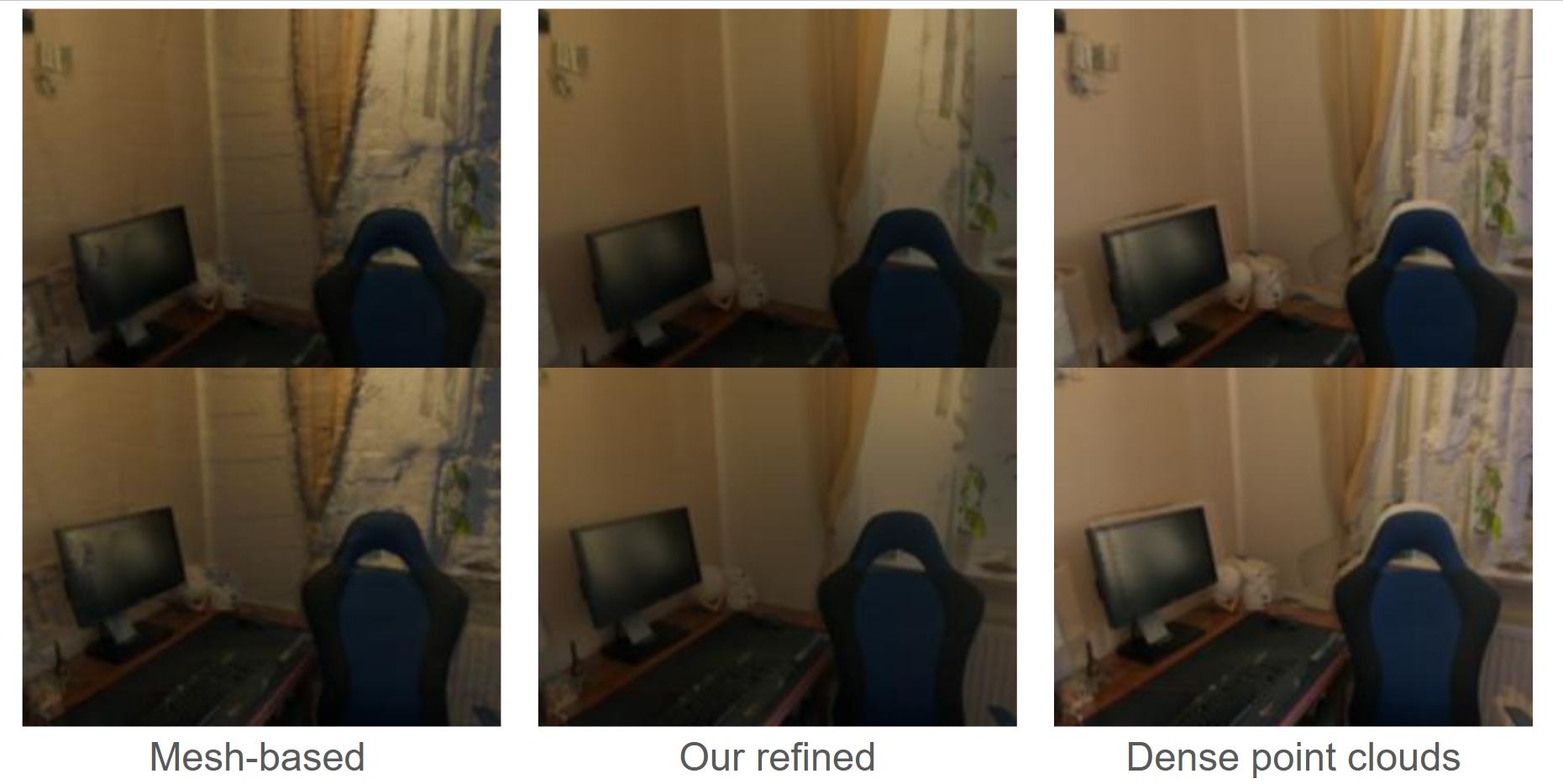}
  \caption{%
        We compare the direct mesh-based filter approach, our method, and a filter based on high-fidelity point clouds. 
  }
  \label{fig:MeshComp}
\end{figure}

For the metrics section, all data was gathered using the mobilenetv4\_conv\_medium model, with cascaded anisotropic diffusion running with 10 steps at 32×32 resolution, 15 steps at 64×64, 25 steps at 128×128, 30 steps at 256×256, and then up-sampled to 512×512 resolutions.
Quantitative results from these benchmarks are summarized in \cref{tab:relighting_metrics}, accompanied by representative visual comparisons illustrated in \cref{fig:meshRefine}, \cref{fig:meshRelight2}, \cref{fig:MeshComp}. These evaluations collectively underscore our approach's robustness, accuracy, and practical applicability for real-time MR relighting tasks.

\begin{table}[h]
\caption{Evaluation metrics for relighting methods on mesh ID 47333462 with 596 images in the ARKitScenes dataset. Lower LPIPS and higher PSNR values indicate better performance. Bold indicates the better (smaller) value for each metric in each benchmark.}
\label{tab:relighting_metrics}
\centering
\begin{tabular}{lcc}
\toprule
Benchmark & ARKitMesh & Ours \\
\midrule
Mesh Error Correction \\
-LPIPS & 0.1322 & \textbf{0.1078} $18\% \downarrow$ \\
-PSNR & 12.7437 & \textbf{12.8100} $0.5\% \uparrow$  \\
\midrule
Multi-Lighting Consistency \\
-LPIPS & 0.4310 & \textbf{0.4120}  $4\% \downarrow$ \\
-PSNR & \textbf{1.7916} & 1.7672 $1.4\% \downarrow$  \\
\midrule
Relighting Fidelity \\
-LPIPS & 0.1391 & \textbf{0.1200}  $14\% \downarrow$  \\
-PSNR & \textbf{15.7850} & 15.7437  $0.3\% \downarrow$  \\
\bottomrule
\end{tabular}
\end{table}

\subsection{Ablations}
In this section, we build upon the analyses presented by Metzger et al \cite{Metzger_2023_CVPR}. in their training details by specifically investigating key components unique to our inference-time techniques.

\textbf{Effectiveness of the Cascaded Diffusion Approach}

\begin{figure}[]
  \centering 
  \includegraphics[width=\linewidth, alt={A figure showing comparions of using 50 steps of cascaded diffusion, 50 steps of naive diffusion, and 1000 steps of naive diffusion. }]{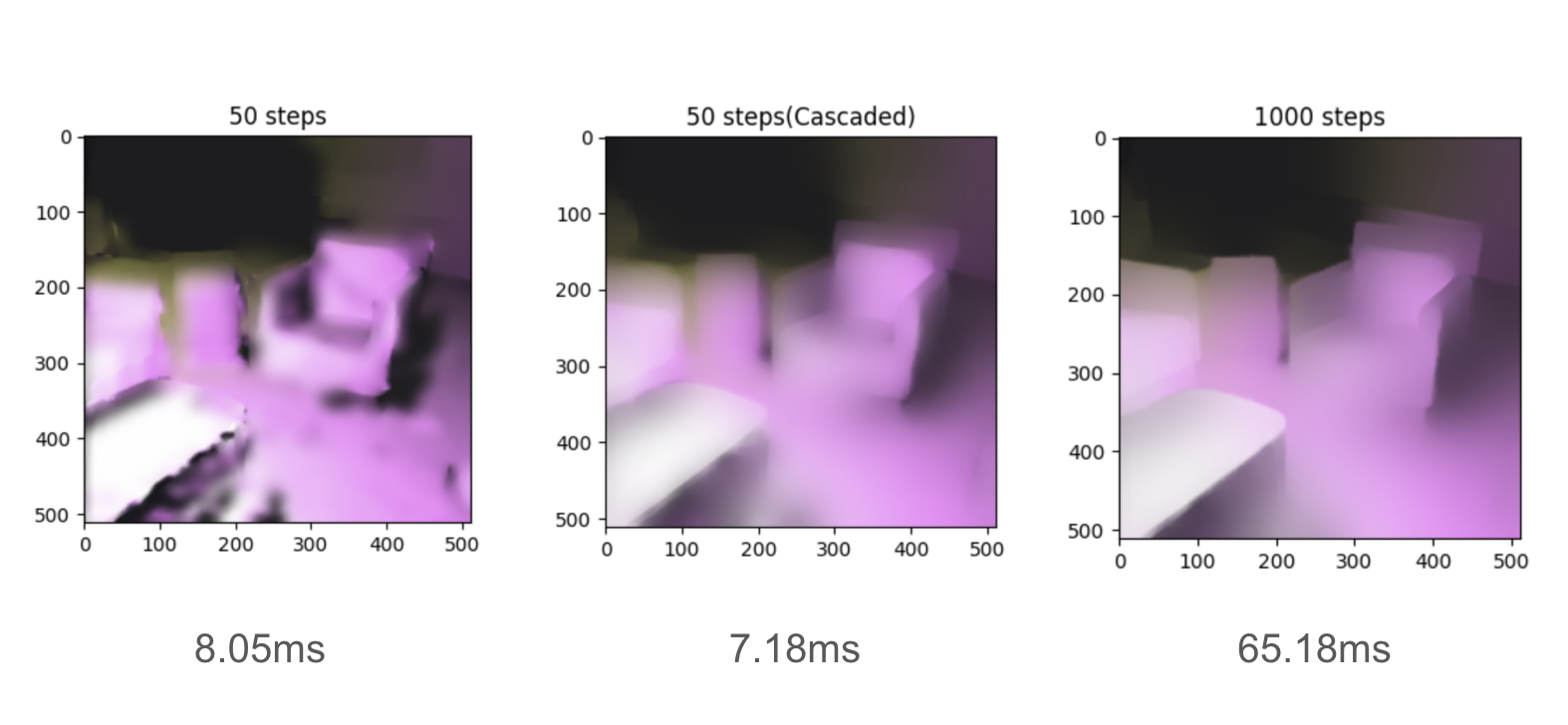}
  \caption{%
        Cascaded diffusion can quickly propagate gradients for high visual quality at a fraction of the time to run.
  }
  \label{fig:cascaded_comp}
\end{figure}

The cascaded diffusion strategy is engineered primarily to reduce the number of diffusion iterations, thereby achieving real-time performance suitable for deployment on edge devices. Here, we rigorously assess its effectiveness and quantify the speed-up it offers. We employ the medium-sized MobileNet model, applying a cascaded diffusion sequence comprising 5 steps at 32×32 resolution, 10 steps at 64×64, 15 steps at 128×128, 20 steps at 256×256, and then up-sampled to 512×512 resolutions. This configuration totals 50 diffusion steps distributed across multiple scales. We compare this cascaded strategy against two baseline scenarios: 50 and 1000 diffusion steps, all executed solely at the resolution of 256×256 and up-sampled to 512x512. We also measure the runtime performance for each scenarios. The result is shown in \cref{fig:cascaded_comp}.

\textbf{Separate Rendering Pass for Shadows}

\begin{figure}[]
  \centering 
  \includegraphics[width=\linewidth, alt={A figure comparing results with and without separate shadow pass.}]{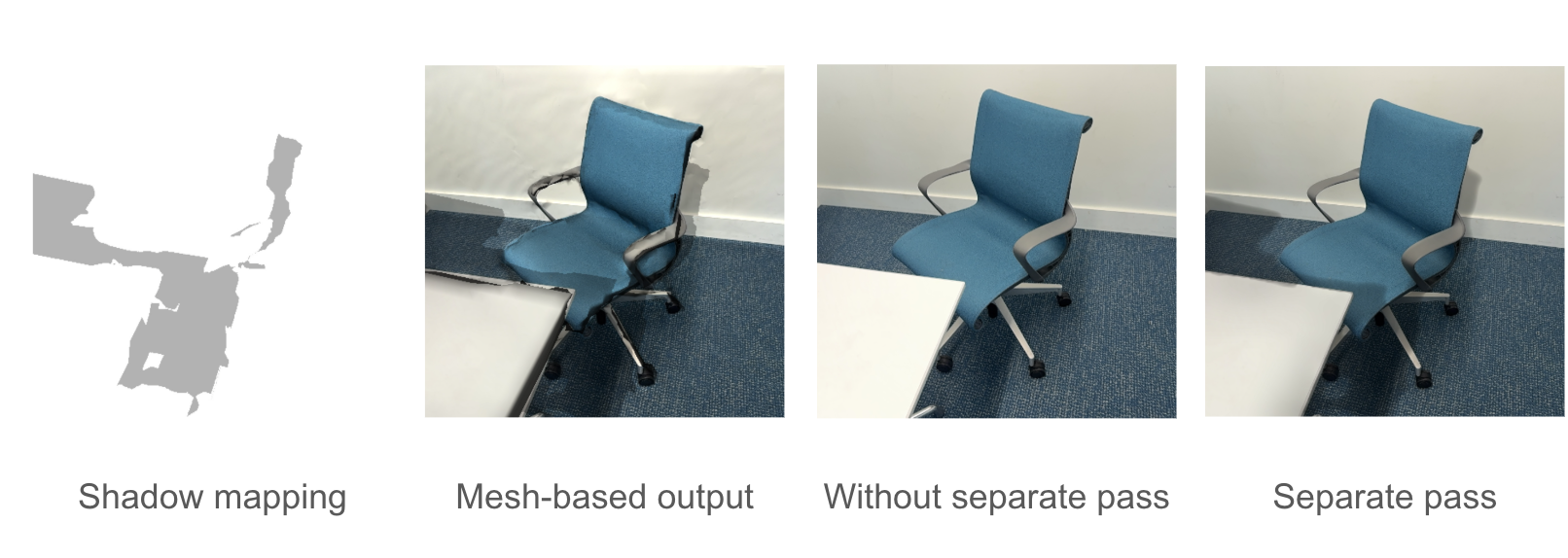}
  \caption{%
       A naive diffusion approach can fail when the shadow is cast onto simple flat surfaces such as walls or floors. A separated pass for shadow rendering can effectively retain the overall shape.
  }
  \label{fig:shadow_comp}
\end{figure}

We further examine the impact of conducting the diffusion process directly on relighting images that include pre-applied shadows, compared to executing the diffusion on shadow-free relighting images and subsequently applying shadows in a separate rendering pass. In \cref{fig:shadow_comp}, we illustrate how directly diffusing images with embedded shadows significantly deteriorates shadow fidelity, primarily due to lack of strict geometric confinement. Conversely, applying shadows separately and refining through targeted shadow map iterations effectively preserves and enhances shadow quality, mitigating the degradation introduced by direct diffusion.

\subsection{Use Case Demonstration}
To demonstrate our approach in a practical mixed reality context, we created a Unity-based demo targeted primarily at handheld devices, such as iPads, suitable for real-estate tour scenarios. Our demo leverages ARKit’s scene reconstruction capabilities to swiftly scan the environment, allowing users to define the positions and orientations of windows through intuitive rectangular area markers. Subsequently, users can interactively visualize different lighting scenarios across various times of the day using a slider interface. This capability remains effective regardless of the actual ambient lighting conditions at runtime, provided the environment is reasonably illuminated.
Additionally, the demo includes scenarios such as cloudy day lighting, offering flat, uniform illumination, and nighttime illumination featuring moonlight effects.

An important extension of our demo addresses custom virtual lighting configurations, highly relevant in practical scenarios such as virtual product demonstrations or personalized interior design planning. Users can intuitively place various virtual light sources, including lamps, candles, or decorative lighting, directly within the real-world environment. The system then dynamically calculates and visualizes realistic lighting interactions and shadows, enabling users to evaluate aesthetic and practical implications without physically altering the actual lighting setups.

Moreover, our demo supports fully dynamic mesh updating, which significantly enhances user interactivity and realism. Users can freely rearrange furniture or other elements within the room and immediately observe the updated lighting effects. This dynamic capability allows real-time assessment of various spatial arrangements, significantly improving user experience by ensuring seamless interaction and immediate visual feedback.

\begin{figure}[h]
  \centering 
  \includegraphics[width=\linewidth, alt={A figure showing our demo app, that helps users to visualize different times of the day.}]{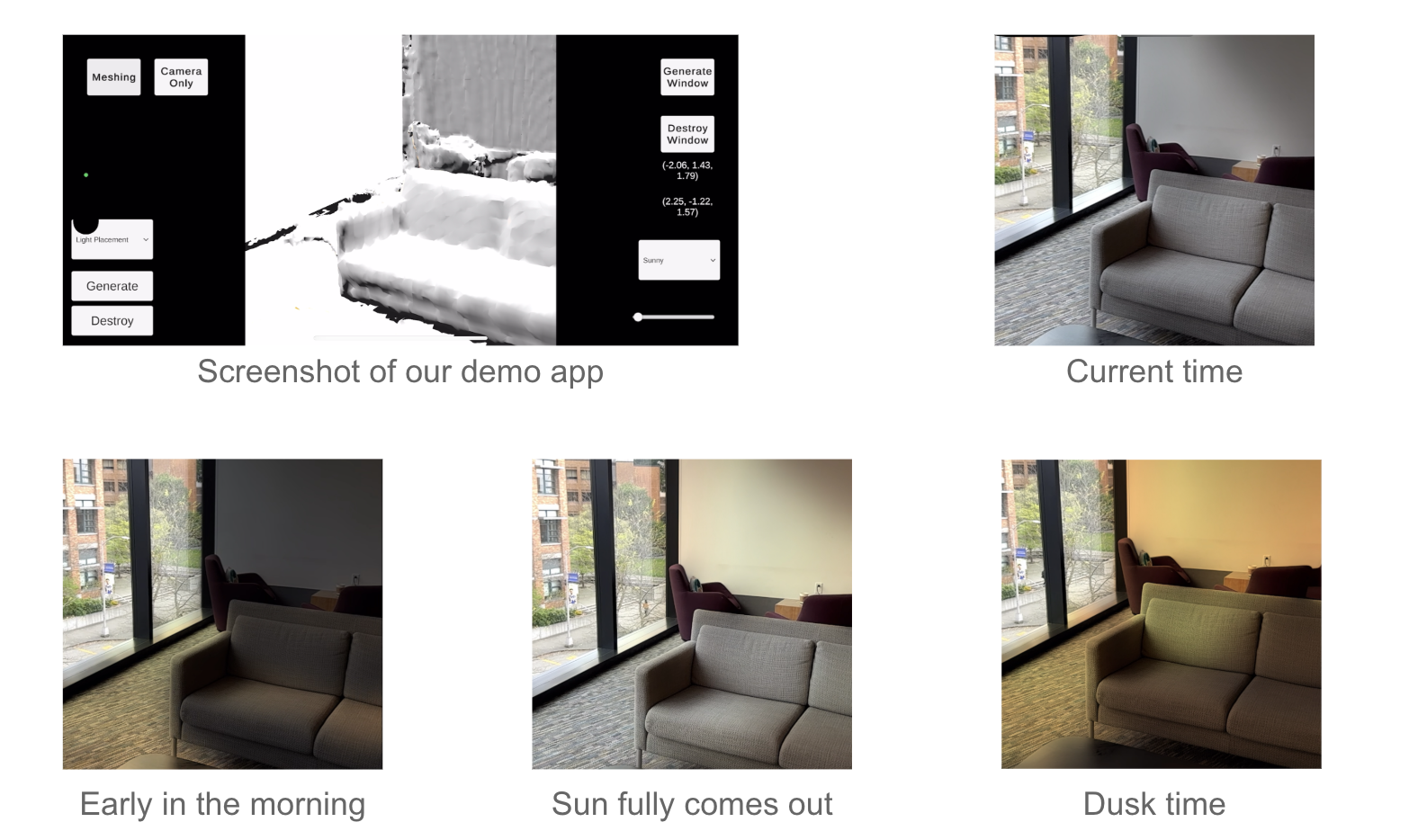}
  \caption{%
        Our demo running in real-time at 50 fps on an iPhone 16 Pro, scanning the environment. We map the positions of the real windows into our demo in order to visualize the lighting more accurately. The top-right image shows what the actual environment looks like. The bottom three images show different times of the day. All images are captured at the same time of the day.
  }
  \label{fig:demo}
\end{figure}

\begin{figure}[h]
  \centering 
  \includegraphics[width=\linewidth, alt={A figure showing our demo app, that helps users to visualize different times of the day.}]{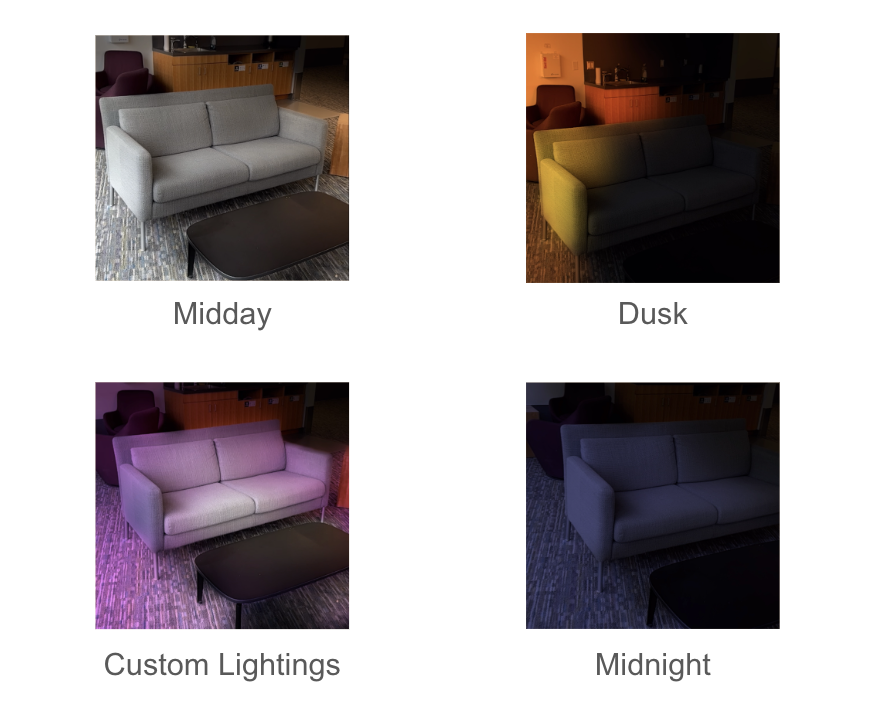}
  \caption{%
        Examples of the room tour demo visualizing different times of the day, and manually placing lights around the room.
  }
  \label{fig:demo2}
\end{figure}

\section{Conclusions}

In this work, we introduced Hybrelighter, a hybrid scene reconstruction and mesh-based filtering approach leveraging guided deep anisotropic diffusion for real-time relighting on edge devices. Our approach successfully combines the computational efficiency of 2D image filters with the depth and geometric accuracy achievable from scene reconstruction capabilities commonly integrated into mixed reality hardware. To address visual artifacts resulting from low-fidelity meshes typically generated by these devices, we demonstrated how guided deep anisotropic diffusion can effectively refine relighting outcomes. Furthermore, we validated the transferability of anisotropic diffusion models originally trained for guided depth super-resolution tasks to our scenario with minimal adjustments.

We also proposed various improvements, including cascaded anisotropic diffusion and a dedicated shadow processing pass, significantly boosting performance and preserving critical visual details such as shadows. Experimental evaluations against direct mesh-based methods confirm that our approach produces visually superior results, capturing essential high-frequency image features, like edges, more effectively. Additionally, comparisons against high-fidelity dense point cloud renderings underline our method's accuracy and realism.

Looking ahead, several promising avenues could further enhance our approach. A key limitation identified in our current pipeline is the absence of a unified framework capable of consistently distinguishing between shadows and geometry-related errors introduced by scanning inaccuracies. Resolving this ambiguity remains a challenging, ill-posed problem. Future research could explore additional learning-based strategies, as well as complementary learning-free methods, to reliably disentangle shadow regions from reconstruction artifacts, ultimately leading to even more robust and visually consistent real-time relighting solutions for mixed reality applications.

\section*{Figure Credits}
\label{sec:figure_credits}
\Cref{fig:orig_paper} is a creation from \cite{Metzger_2023_CVPR}

\acknowledgments{%
	The authors wish to thank A, B, and C.
  This work was supported in part by a grant from XYZ (\# 12345-67890).%
}

\bibliographystyle{abbrv-doi-hyperref}

\bibliography{MR-relighting-review-submission}

\end{document}